\documentclass[a4paper]{jpconf}
\usepackage{graphicx}
\usepackage{amssymb}
\usepackage{color}
\usepackage{amsmath}
\newcommand{\ra}{\rangle}
\newcommand{\la}{\langle}
\newcommand{\be}{\begin{equation}}
\newcommand{\ee}{\end{equation}}
\newcommand{\bea}{\begin{eqnarray}}
\newcommand{\eea}{\end{eqnarray}}
\newcommand{\bd}{\begin{displaymath}}
\newcommand{\ed}{\end{displaymath}}

\graphicspath{{figv3/}} 

\begin{document}

\title{Topological properties of a Valence-Bond-Solid}
\author{Hui Shao$^1$, Wenan Guo $^{1, 2, *}$ and Anders W. Sandvik$^{3,\dagger}$}

\address{$^1$ Department of Physics, Beijing Normal University, Beijing 100875, China}
\address{$^2$ State Key Laboratory of Theoretical Physics, Institute of Theoretical
Physics, Chinese Academy of Science, Beijing 100190, China}

\address{$^3$ Department of Physics, Boston University, 590 Commonwealth Avenue, Boston, Massachusetts 02215, USA}
\ead{$^*$waguo@bnu.edu.cn,~~$^\dagger$sandvik@bu.edu}
\begin{abstract}
We present a projector quantum Monte Carlo study of the topological properties of the valence-bond-solid ground state 
in the $J$-$Q_3$ spin model on the square lattice. The winding number is a topological number counting the number of domain walls in 
the system and is a good quantum number in the thermodynamic limit. We study the finite-size behavior and obtain the domain wall energy 
density for a topological nontrivial valence-bond-solid state.
\end{abstract}

\section{Introduction}

While topological order has mainly been discussed in the context of exotic states such as quantum spin liquids \cite{spinliquid,Tang}, similar 
topological quantum numbers can also arise in systems with long-range order.  The simplest example of a topological quantum number is the winding 
number of a dimer model \cite{RK}. In the classical close-packed dimer model on the square lattice (as an example of a simple bipartite lattice), 
a winding number $W = (w_x, w_y )$ can be defined by assigning a directionality (arrow) from sublattice A to sublattice B for each dimer. 
Superimposing any such configuration of arrows onto two reference configurations (conventionally one with the dimers forming horizontal or vertical 
columns), closed loops are formed and the winding numbers correspond to the total $x$ and $y$ currents normalized by the system length $L$ \cite{RK}. 

In the quantum dimer model (QDM), the classical dimer configurations are the basis states. For $S = 1/2$ quantum spins on a bipartite lattice, any total 
spin singlet can be expanded in Valence-bond states (VBs) with positive-defined expansion coefficients. The VBs are dimers with the added 
spin-singlet internal structure and the configurations can again be classified according to the winding number \cite{bonesteel89}. Since the Hamiltonian 
of the QDM contains only local operators that have no matrix element between states of different winding numbers, the Hilbert space can be divided 
into different winding number sectors, i.e., the topological winding number is a good quantum number. However, for general VB states, longer bonds 
are introduced, which makes the bond operators in the Hamiltonian non-local. For finite size systems, the winding number can be changed if bonds
of lengths exceeding $1/4$ of the system length appear. Therefore, the winding number can only be conserved in the thermodynamic limit.

\begin{figure}[h]
\begin{center}
\includegraphics[width=0.12\linewidth]{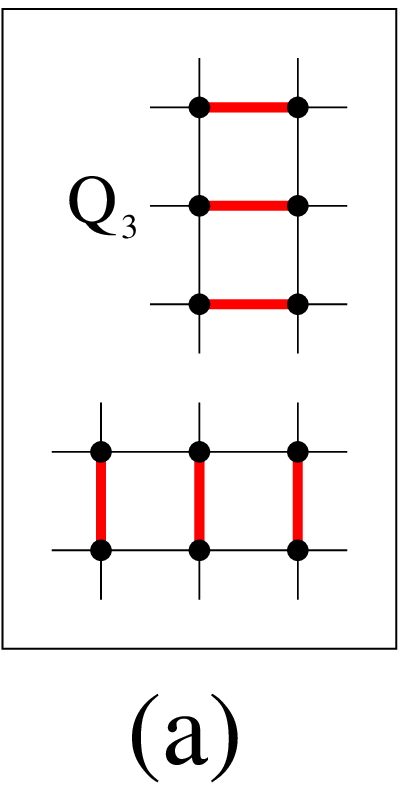}~~~~~~~
\includegraphics[width=0.17\linewidth]{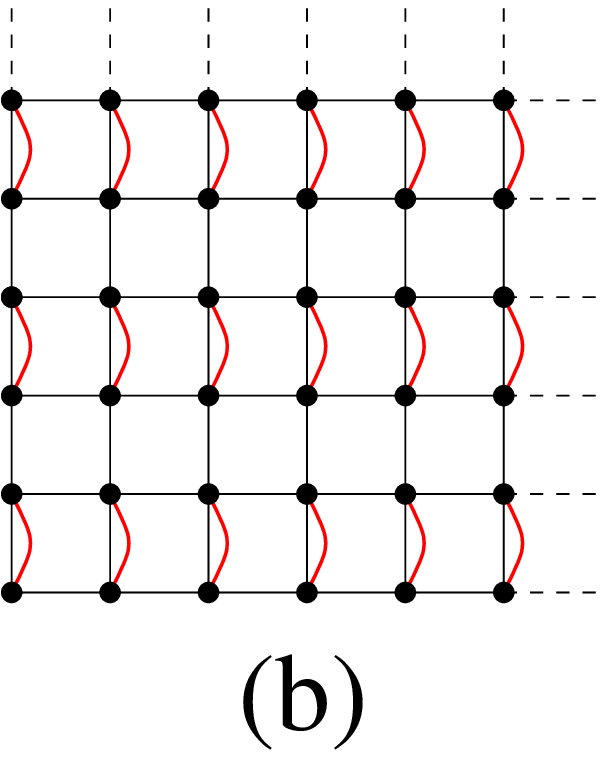}~~~~~
\includegraphics[width=0.17\linewidth]{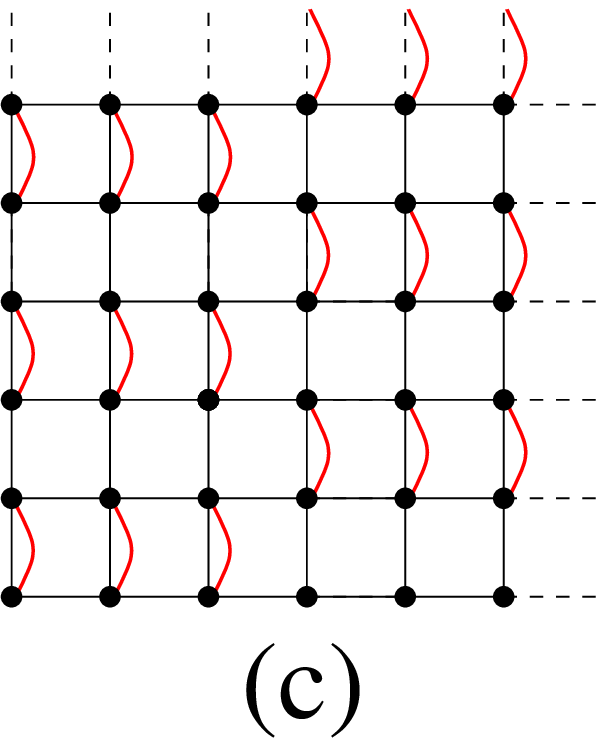}
\end{center}
\vskip-5mm
\caption{\label{VBS} Illustrations of (a) the $Q_3$ terms, (b) the columnar VBS state, and (c) 
a VBS state with winding number $W=(1,0)$ (on a periodic lattice).}
\end{figure}

To investigate the emergent conserved winding number in an $S = 1/2$ spin system, we here study the $J$-$Q_3$ Model on the square 
lattice \cite{JQ}. The Hamiltonian of the model is
\be
H=-J\sum_{\la ij \ra} C_{ij} -Q_3 \sum_{\la ij,kl,mn\ra} C_{ij}C_{kl}C_{mn}
\label{H}
\ee
with $C_{ij}=\frac{1}{4}-{\bf S}_i\cdot {\bf S}_j$ recognized as a singlet projector. The $J$ term of $H$ is the standard Heisenberg exchange and
the $Q_3$ terms have the singlet projectors arranged in columns, as shown in Fig.~\ref{VBS}(a).  This model hosts a columnar valence-bond solid (VBS) 
ground state when  $Q_3/J$ is sufficiently large \cite{lou09}. Ignoring quantum fluctuations, the VBS ground state is sketched as an ordered VBS
breaking the $Z_4$ symmetry of the Hamiltonian, as shown in Fig. \ref{VBS}(b). As previously studied in the context of QDMs \cite{papanikolaou07},
a VBS with domain walls on a periodic lattice corresponds to a non-zero winding number---a case of $W=(1,0)$ is shown in Fig. \ref{VBS}(c).

In this paper, we consider the pure $Q_3$ model without any $J$ term to study the VBS ground state.
We will demonstrate explicitly, using projector quantum Monte Carlo (PQMC) calculations, that there is an emergent 
topological quantum number (a winding number) associated with the VBS order in the thermodynamic limit, while in a finite 
system the winding number is not conserved. In this case the topological sectors correspond to different types of domain walls, 
which are unstable (non-conserved) in finite systems but become stable in the thermodynamic limit.

\section{Projector Quantum Monte Carlo method}

The model with $Q_3>0$ in (\ref{H}) does not have sign problems. We thus take advantage of the PQMC method with efficient loop update 
algorithms \cite{AEloop,loopEL,loopE} to study the model. The method is based on applying a high power of the Hamiltonian to 
a trial state $|\psi_0\ra$,
\be
|\psi_m\ra = (-H)^m |\psi_0\ra.
\ee
For sufficient large $m$, $|\psi_m\ra$ approaches the ground state $|0\ra$ of the model. As the ground state $|0\ra$ of a 
bipartite $J$-$Q_3$ model is guaranteed to be a total spin singlet, it is particularly convenient to use a trial state expanded 
in the VB basis $|V_r\ra$ in the singlet sector. Then only the singlet sector is considered from the outset, leading to
faster convergence with $m$.

The energy of the projected state can be calculated in the following way \cite{sandvik05}
\be
E=\frac{\la N|H |\psi_m \ra}{\la N|\psi_m \ra}= -\frac{\la N|(-H) (-H)^m |\psi_0\ra }{\la N | (-H)^m |\psi_0\ra},
\ee
where $|N\ra$ is one of the N\'eel states in the $z$-spin basis, which has equal overlap with all VBs. 

By writing $(-H)^m$ as a sum over  all possible strings of the individual $Q_3$ terms in Eq. (\ref{H}), the above expression is 
evaluated by implementing importance sampling of the operator strings with the efficient loop algorithm \cite{loopEL,loopE}. 
The implementation of the algorithm to the $J$-$Q_3$ Hamiltonian has been described briefly in \cite{JQpqmc}. In the end,
the energy $E$ is calculated as 
$E=-\la n_d +n_f/2\ra$
with $n_f$ the number of bond flips and $n_d$ the number of diagonal operations. 

\section{Results}

We chose a basis state $|\psi_0\ra=|V_r\ra$ as the trial state and consider the cases that $|V_r\ra$ in different winding number sectors 
$W$. The detailed definition of the winding number of a VB state can be found, e.g., in Refs.~\cite{RK,Tang}. In the PQMC simulations,
besides the energy, we also sample the probability, $P(W)$, of a projected state in the topological sector $W=(w_x, w_y)$. This 
is done by calculating the winding number of each projected VBs $P_k|V_r\ra$, with $P_k$ a operator string with length $m$ generated 
in the MC processes.

In the case that  the trial state is a VBs  in the winding number sector $W=(0,0)$, the ground columnar VBS state will be projected out 
quickly, i.e., within a small $m/N$, which can be defined as a projecting "time" (closely related to imaginary time \cite{liu13}). This 
is indicated by the convergence of the ground state energy density $e_0(L)$ for a system with linear size $L$.  

Now turn to the cases that the trial state is in the nontrivial winding number sector $W\ne (0,0)$. For small systems, the columnar VBS state 
is again projected out after some projection time $m/N$. This is indicated by the convergence of the energy density $e(L)$ to the ground state 
value $e_0(L)$. Meanwhile, the probability $P(W)$ decreases (to $0$ for $L \to \infty$). However, as the system size increases, the projection 
time $m/N$ needs to grow as well in order for the ground state to be obtained. In the thermodynamic limit, we expect that the system will stay in 
the sector with the initial winding number $W$, and then it is also plausible that the energy of the system will converges to a value $e_W>e_0$ 
corresponding to the lowest excited state within the sector $W$.

\begin{figure}[h]
\begin{minipage}{7cm}
\begin{center}
\includegraphics[width=0.9\linewidth]{L96}
\end{center}
\caption{\label{PE} The probability $P(W)$ and the energy density $e_W(L)$ as functions of time $m/N$ (projector power
rescaled by the system volume).}
\end{minipage}
\hspace{0.2cm}
\begin{minipage}{8.5cm}
\includegraphics[width=8.5cm]{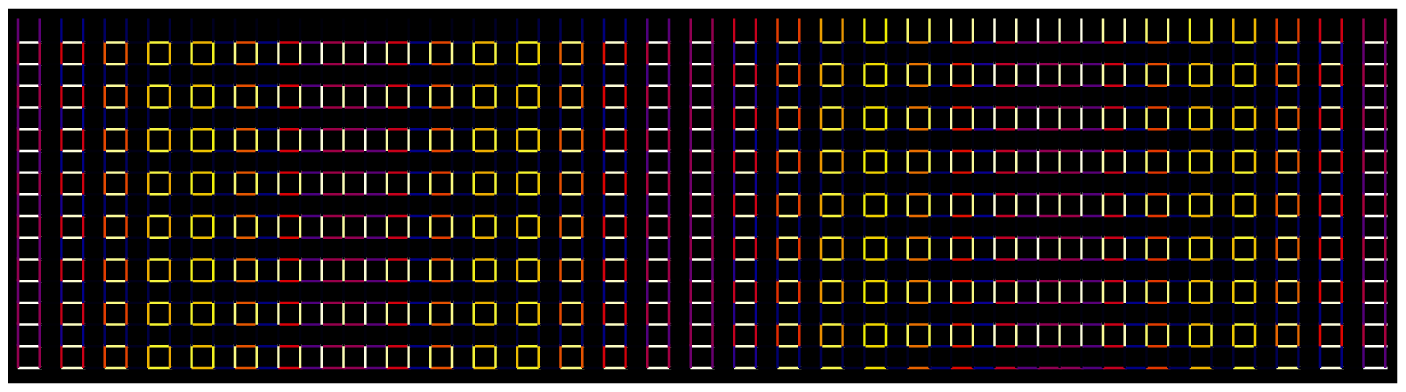}
\includegraphics[width=8.5cm]{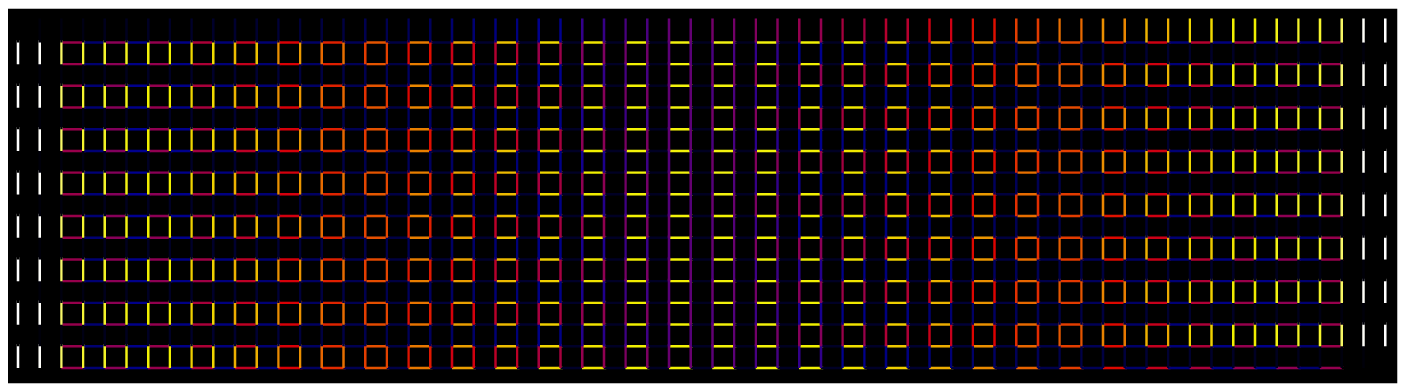}
\caption{\label{PiDW} Snapshots of $\la B_{\alpha}({\bf r})\ra$ for a periodic system with winding number $W=(1,0)$, in which
a $2 \pi$ domain wall (four separate $\pi/2$ domain walls) is formed (upper panel) and for an open system with appropriate boundary 
conditions in which a $\pi$ domain wall is forced (lower panel).}
\end{minipage}

\end{figure}

To demonstrate such behavior, we introduce the energy density $e_W(L)$ of states in the winding number sector $W$, which is obtained by only 
sampling those states in the sector $W$. Figure \ref{PE} shows the "time evolution" of the probability $P(W)$ for the system staying in the 
original winding number sector $W=(0,0), W=(1,0),$ and $W=(2,0)$ as a function of $m/L^2$, respectively, for a system with linear size 
$L=96$ (lower panel). The corresponding energy density $e_W(L)$ converges to the values which are higher than $e_0(L)$, if $W \ne 0$ (upper panel). 

We now study the reason of the energy gap between a system in a nontrivial topological sector and in the $W=0$ ground state.
It is well known that the ground state of the $J$-$Q_3$ is the columnar VBS. The VBS state 
can be detected by the columnar VBS order parameter, which is defined by the operators
\be
D_x=\frac{1}{N}\sum_{x,y}B_{\hat{x}}(x,y)(-1)^x,~~~~~
D_y=\frac{1}{N}\sum_{x,y}B_{\hat{y}}(x,y)(-1)^y. 
\ee
where the dimer operator $B_{\alpha}$ is used:
\be
B_{\alpha}({\bf r})= {\bf S}({\bf r}) \cdot {\bf S}({\bf r}+\alpha)
\ee
where $\alpha=\hat{x},\hat{y}$ denotes the lattice unit vector in the two directions.
In a columnar VBS, either $D_x$ or $D_y$ has a nonzero expectation value.

However, if a state bears a nontrivial winding number, the lattice translational symmetry with period two is also broken.  As an example, 
Fig.~\ref{PiDW} (upper panel) shows a snapshot of the bond configuration $\la B_{\alpha}({\bf r})\ra$ for a projected state with winding number $W=(1,0)$.
To describe such a state, it is useful to define the local order parameters for the VBS state. Consider the case that there is still translational 
invariant VBS with period two along one axis of the square lattice, say $y$ axis. The one-dimensional local $x$ and $y$ order parameters as a 
function of the $x$ coordinate are defined \cite{FS} as follows:
\bea
D_x(x)&=&[\langle B_{\hat{x}}(x,y)\rangle-\frac{1}{2}\langle B_{\hat{x}}(x-1,y)\rangle-\frac{1}{2}\langle B_{\hat{x}}(x+1,y)\rangle](-1)^x,\\
D_y(x)&=&[\la B_{\hat{y}}(x,y)\ra-\la B_{\hat{y}}(x,y+1)\ra](-1)^y.
\eea
These two local order parameters are independent of the $y$ coordinate. Following theoretical expectations \cite{levin04},
The VBS angle $\theta(x)$ can also be defined \cite{FS},
\be
\theta(x)={\rm atan}\left [\frac{D_y(x)+D_y(x+1)}{2D_x(x)} \right],
\ee
such that $\theta=0$ and $\theta=\pi$ for a fully $x$ or $y$ oriented VBS order, respectively.

\begin{figure}[h]
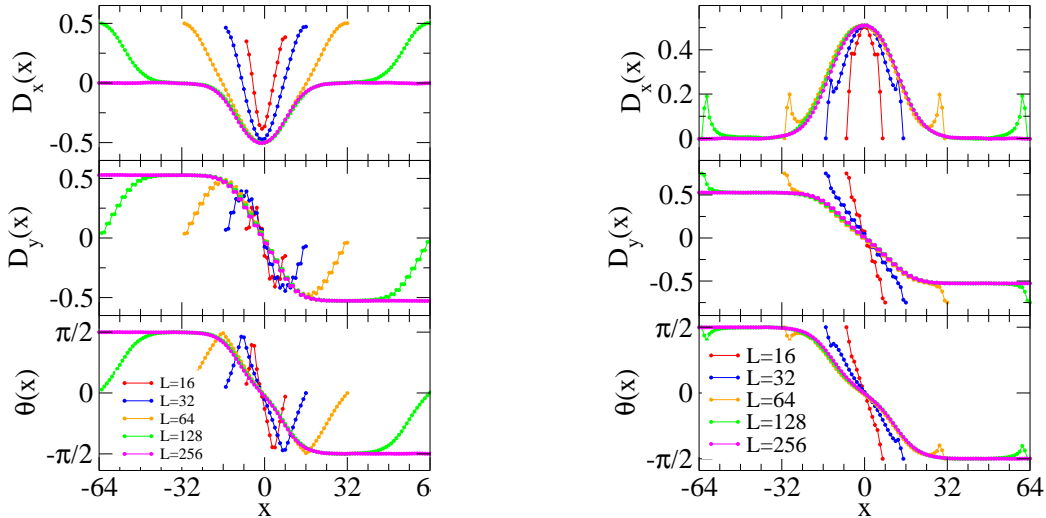

\vspace{1cm}
\begin{center}
\includegraphics[width=0.36\linewidth]{D-10}\hspace{2cm}
\includegraphics[width=0.36\linewidth]{D-single}
\end{center}
\vskip-4mm
\caption{\label{DW} The local order parameters $D_x(x)$, $D_y(x)$ and the VBS angle $\theta(x)$ for projected states in the 
topological sector $W=(1,0)$ with periodic boundaries (left) and for the states with a $\pi$ domain wall (right) in a system with asymmetric 
open boundaries.}
\end{figure}

To understand the energy gap between system in an nontrivial topological sector and in the $W=0$ ground state, we first set asymmetric open 
boundary conditions on a $L \times L$ cylindrical lattice, where the $x$-direction boundaries have been modified. By removing $Q_3$ terms with 
vertical bonds closest to the left edge on every even row and closest to the right edge on every odd row, we obtain a state with forced VBS 
angle $\theta(0)=\pi/2$ and $\theta(L)=-\pi/2$. A snapshot of the bond strength $\la B_{\alpha}({\bf r})\ra$ for the state projected out from 
simple trial state with a corresponding domain wall is shown in Fig.~\ref{PiDW} (lower panel). More properties of the projected state are shown 
in Fig.~\ref{DW} (right), where the two local order parameters and the VBS angle are plotted as functions of $x$ coordinate and various 
system sizes $L$. In this case, the local order parameter $D_x(x)$ is maximized in the center and goes to zero at boundaries; $D_y(x)$ is 
about $0.75$ at the left boundary, but gradually changes to $-0.75$ at the right boundary. The VBS angle $\theta(x)$ changes 
from $\pi/2$ to $-\pi/2$. Clearly a domain wall is formed, as expected.  According to the total change of the VBS angle, we define this
as a $\pi$ domain wall (which consist of two $\pi/2$ domain walls). 

\begin{figure}[h]
\begin{center}
\includegraphics[width=0.43\linewidth]{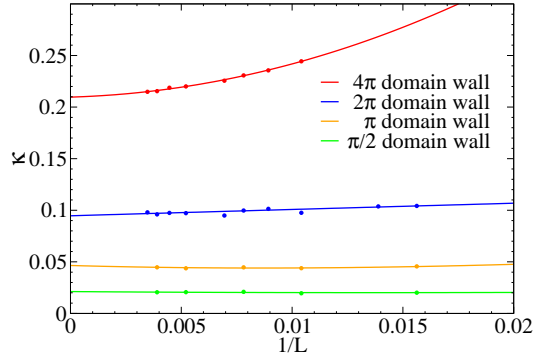}
\end{center}
\vskip-5mm
\caption{\label{dwe} Finite-size scaling analysis of the domain wall energies. The curves are fitted polynomials, used
to extrapolate the energies listed in Table \ref{dwetab}.}
\end{figure}

The energy density of the state with $\pi$ domain wall $e_{\pi}(L)$ can be calculated as described above. The resulting $e_{\pi}(L)$ is larger than the 
ground state energy density $e_0(L)$ of a $L \times L$ cylindrical lattice with symmetric boundary modification (i.e., forcing a VBS with no domain
wall). The energy difference is entirely due to the presence of the domain wall. We thus define the domain wall energy per unit length as
\be
\kappa_{\pi}(L) =[e_{\pi}(L)-e_0(L)]L
\label{kappadef}
\ee
As shown in Fig. \ref{dwe}, when system size goes to infinity, the domain wall energy per unit length 
converge to a constant $\kappa_{\pi}$, which is estimated as 0.434(9), as listed in Table \ref{dwetab}. 

Similarly, a $\pi/2$ domain wall can realized by 
setting asymmetric open boundary conditions forcing horizontal and vertical bond patterns at the edges. The energy difference to $e_0(L)$ 
multiplied by the system size again converges to a constant when system size $L$ tends to infinity (see Fig.~\ref{dwe}). The constant is thus identified 
as the domain wall energy $\kappa_{\pi/2}$ for the $\pi/2$ domain wall with the estimated value being a half of the $\pi$ domain wall energy, 
as listed in Tab.~\ref{dwetab}. 

We now return to systems with periodic boundaries. In the case of large enough system, the projected energy eigenstate is still in the winding number 
sector $W$ of the initial state. Take the $W=(1,0)$ case as an example. Figure \ref{DW} (left panel) shows the local order parameters and VBS angel 
as functions of the $x$ coordinate. $D_x(x)$ has a minimum in the center and maximums at the left and right sides, while $D_y(x)$ is almost zero at 
the edges  gradually grows to a maximum of $0.5$, then decays to a minimum about $-0.5$, and finally returns to $0$. As the result, the VBS 
angle $\theta(x)$ changes from $0$ to $\pi/2$, then to $-\pi/2$, and back to $0$. Domain walls are clearly formed here and their dynamics in a 
large system is very slow and practically locked in to their locations in the initial state. According to the total change of the VBS angle 
we define this as a $2\pi$ domain wall (which can be regarded as four elementary $\pi/2$ domain walls). A finite-size scaling analysis of the 
energy density gap to the ground state ($W=0$) $e_W(L)-e_0(L)$ multiplied by the system size $L$ is shown in Fig.~\ref{dwe}. Clearly, the scaled 
gap converges to a constant, which can be understood as the $2\pi$ domain wall energy per unit length $\kappa_{2\pi}$.  The estimated $\kappa_{2\pi}$ 
is indeed twice that of the $\pi$ domain wall, as seen in Table \ref{dwetab}.

We further calculated the energy density gap $e_W(L)-e_0(L)$ from the projected state in the winding number sector $W=(2,0)$ to the ground state for several
system sizes.  The local order parameters and VBS angles as functions of the $x$-direction shows that there is a $4\pi$ domain wall. Again the energy gap
multiplied by the system size converges to a constant, which is the $4 \pi$ domain wall energy. The estimated domain wall energy is $0.205(9)$, which is 
about four times of the $\pi$ domain wall energy.

\begin{table}[h]
\begin{center}
\begin{tabular}{c|c|c|c|c}
\hline
domain wall & $\pi/2$ &  $\pi$ & $2 \pi$ & $4 \pi$  \\
\hline 
$m/N$       &6        &    6   &  4      &   3          \\
\hline
$\kappa$    &0.0208(5) &0.0434(9) &0.093(9)  &0.205(9)   \\
\hline
\end{tabular}
\end{center}
\vskip-3mm
\caption{\label{dwetab} Domain wall energy per unit length for various domain walls. $m/N$ is the projection ``time'' at which the energy 
is calculated in each case.}
\end{table}

\section{Summary and Conclusions}

Using a ground-state projector QMC method in the VB basis, we have studied projected states in various winding number sectors of the 
$J$-$Q_3$ model on the square lattice. We showed that the projected state stays in the winding number
sector of the trial state when system size goes to infinity. The energy of the states in a nontrivial winding number sector have gaps to the VBS
ground state in the $W=(0,0)$ sector, due to the presence of domain walls (the winding number counting these domain walls). Such domain walls 
are stable only for infinite size, thus, the winding number is an emergent quantum number in the thermodynamic limit. 

\ack  This research was supported by the NSFC under Grant No.~11175018 (WG) and by the NSF under Grants No.~DMR-1104708, PHY-1211284 and
DMR-1410126 (AWS). SH gratefully acknowledges support from the organization committee of the CCP2014. WG would like to thank Boston University's 
Condensed Matter Theory Visitors program.

\end{document}